\documentclass[pdflatex,sn-mathphys-num]{sn-jnl}% Math and Physical Sciences Numbered Reference Style

\usepackage{amssymb}
\usepackage{amsmath}
\usepackage{amsfonts}
\usepackage{textcomp}
\usepackage{xcolor}
\usepackage{url}
\usepackage{booktabs}
\usepackage{graphicx}

\usepackage{graphicx}%
\usepackage{multirow}%
\usepackage{amsmath,amssymb,amsfonts}%
\usepackage{amsthm}%
\usepackage{mathrsfs}%
\usepackage[title]{appendix}%
\usepackage{xcolor}%
\usepackage{textcomp}%
\usepackage{manyfoot}%
\usepackage{booktabs}%
\usepackage{algorithm}%
\usepackage{algorithmicx}%
\usepackage{algpseudocode}%
\usepackage{listings}%

\begin{document}

\title[Towards Human-AI Collaboration System for the Detection of Invasive Ductal Carcinoma in Histopathology Images]{Towards Human-AI Collaboration System for the Detection of Invasive Ductal Carcinoma in Histopathology Images}


\author[1]{\fnm{Shuo} \sur{Han}}\email{sh1276@exeter.ac.uk}

\author*[1,2]{\fnm{Ahmed Karam} \sur{Eldaly}}\email{a.karam-eldaly@exeter.com}

\author[1,3]{\fnm{Solomon Sunday} \sur{Oyelere}}\email{s.oyelere@exeter.ac.uk}

\affil*[1]{\orgdiv{Department of Computer Science}, \orgname{University of Exeter}, \orgaddress{\city{Exeter}, \postcode{EX4 4QF}, \country{United Kingdom}}}

\affil[2]{\orgdiv{Centre for Medical Image Computing}, \orgname{University College London}, \orgaddress{\street{Gower St.}, \city{London}, \postcode{WC1E 6AE}, \country{United Kingdom}}}

\affil[3]{\orgdiv{Research Group on Data, Artificial Intelligence, and Innovations for Digital Transformation}, \orgname{Johannesburg Business School, University of Johannesburg}, \orgaddress{\city{Johannesburg}, \country{South Africa}}}


\abstract{Invasive ductal carcinoma (IDC) is the most prevalent form of breast cancer, and early, accurate diagnosis is critical to improving patient survival rates by guiding treatment decisions. Combining medical expertise with artificial intelligence (AI) holds significant promise for enhancing the precision and efficiency of IDC detection. In this work, we propose a human-in-the-loop (HITL) deep learning system designed to detect IDC in histopathology images. The system begins with an initial diagnosis provided by a high-performance EfficientNetV2S model, offering feedback from AI to the human expert. Medical professionals then review the AI-generated results, correct any misclassified images, and integrate the revised labels into the training dataset, forming a feedback loop from the human back to the AI. This iterative process refines the model's performance over time. The EfficientNetV2S model itself achieves state-of-the-art performance compared to existing methods in the literature, with an overall accuracy of 93.65\%. Incorporating the human-in-the-loop system further improves the model's accuracy using four experimental groups with misclassified images. These results demonstrate the potential of this collaborative approach to enhance AI performance in diagnostic systems. This work contributes to advancing automated, efficient, and highly accurate methods for IDC detection through human-AI collaboration, offering a promising direction for future AI-assisted medical diagnostics.}


\maketitle

\section{Introduction}
\label{sec:introduction}
As reported by the national breast cancer organisation, breast cancer is the second leading cause of death among women. It is expected to account for 30\% of new cancers diagnosed in women by the end of 2024 \cite{NBCBreastCancerFacts}. Studies have shown that early detection and timely treatment of breast cancer can effectively reduce the death rate of breast cancer \cite{Taylore074684}. In the United States, the 5-year relative survival rate is as high as 99\% if breast cancer is detected locally at an early stage \cite{NBCBreastCancerFacts}. Invasive ductal carcinoma (IDC) accounts for approximately 70-80\% of all breast cancers diagnosed \cite{Ozturk2022}. Therefore, the influence of the diagnostic accuracy of IDC on the treatment effect is significant. In recent years, imaging studies have played a crucial role in the diagnosis and treatment of breast cancer. The primary imaging modalities include screen-film mammography (SFM), digital mammography, and magnetic resonance imaging (MRI) \cite{https://doi.org/10.1002/int.22622}. Although each of these methods has its advantages, none can eliminate the possibility of misdiagnosis. Diagnosis based on histopathological images overcomes this weakness, becoming the most effective and indispensable process after a breast biopsy \cite{10143757}. However, analysing histopathology images is a complex and time-consuming process that relies not only on expert expertise but often also on the subjective judgment of the pathologist \cite{10.3389/fgene.2019.00080}. In addition, the increasing number of patients has led to a severe shortage of pathologists \cite{Jawad2023}. As a result, improving the accuracy and efficiency of breast cancer diagnosis based on histopathology images has become the focus of many studies.

With the rapid development of artificial intelligence (AI) in the field of image analysis, computer-aided diagnosis (CAD) systems are widely used in medical image analysis to help medical experts improve diagnostic accuracy and efficiency \cite{SHIRAISHI2011449, eldaly2015bag, basaad2024bert}. Although traditional machine learning-based CAD methods have achieved promising results, the different feature extraction processes limit their performance \cite{HU2018134}. To overcome this weakness, deep learning has gradually gained attention. It has been proven to outperform machine learning methods in the diagnosis of breast cancer as it can automatically extract features without the need for complex feature engineering \cite{Chugh2021}. At the same time, studies have shown that convolutional neural networks (CNN) have shown significant performance in cancer diagnosis \cite{s20164373, 10.1063/5.0171483, Gnanasekaran, Sohail2021, doi:10.1080/21681163.2020.1824685, GONCALVES2022105205, basaad2024bert}. This allows artificial intelligence to be increasingly used in the medical decision-making process, effectively improving the quality and efficiency of decision-making.

While AI has enhanced the accuracy and efficiency of cancer diagnosis, the performance of diagnostic systems does not solely depend on the capabilities of AI or the expertise of medical professionals. The quality of interaction between the two is the most decisive factor \cite{INTROZZI20241131}. AI models still need to overcome certain obstacles and challenges, such as issues with model validation and algorithm interpretability, which prevent humans from fully trusting AI-generated diagnostic results \cite{cancers14215264, mosqueira2023human}. Therefore, it is necessary for medical experts to optimise and refine the models regularly. To address these challenges, this work proposes a deep learning-based human-in-the-loop system aimed at improving the efficiency and accuracy of IDC detection in histopathology images. The core of this system involves continuously training the model through human intervention in the training data, thereby optimising the model's performance. This approach not only enhances diagnostic accuracy but also incrementally improves system performance through accurate human feedback, offering valuable insights into improving the quality of breast cancer detection and diagnosis. The primary contributions of this work are as follows.
\begin{enumerate}
\item We propose a lightweight deep learning framework, EfficientNetV2S, leveraging transfer learning pre-trained on the ImageNet dataset for the detection of invasive ductal carcinoma (IDC) in histopathological images.
\item We evaluate the model’s performance on publicly available datasets, achieving state-of-the-art results when compared to recent methods in the literature.
\item We introduce a human-AI collaboration system based on a human-in-the-loop approach for the detection of IDC in histopathological images. Human experts correct misclassified images and incorporate the revised labels into the training dataset. This iterative process retrains the model, significantly improving its generalisation capability. Experimental results demonstrate that this method effectively enhances the model’s performance.
\end{enumerate}

The remaining sections of the paper are organised as follows. Section \ref{sec:Literature Review} reviews the literature for ductal carcinoma detection. The design and implementation details of the proposed approach are presented in Section \ref{sec:Methodology}. Experimental results are then presented in Section \ref{sec:Results}. Discussions and general insights into the proposed approach, including main limitations and plans for future work are discussed in Section \ref{sec:Discussion}. Finally, conclusions are outlined in Section \ref{sec:Conclusions}.

\section{Related Work}
\label{sec:Literature Review}
This section reviews the literature on breast cancer detection, focusing on histopathology images, as investigated in this work. It also highlights the significance of human involvement in human-AI collaborative systems to enhance the performance of computer-aided diagnosis (CAD) systems.

Breast cancer presents diverse characteristics across various imaging techniques, requiring experts to make informed decisions based on the strengths and limitations of each method \cite{00005792-202105140-00037}. However, non-invasive techniques such as SFM and MRI often struggle to accurately pinpoint cancer locations, which may lead to diagnostic errors. To address this, Hameed et al. \cite{s20164373} advocated for the use of histopathology image analysis for more comprehensive and reliable decision-making. Likewise, Kadhim et al. \cite{idc} demonstrated that histopathology images can support more precise classification, evaluation, and treatment planning in breast cancer diagnosis.

Artificial intelligence (AI) has become an invaluable tool in medical image analysis, with machine learning playing a particularly important role \cite{10.1007/978-981-13-1217-5_73, Naik2023, 9867097, 7312934, eldaly2015bag, eldaly2019bayesian, eldaly2019patch}. Goyal et al. \cite{10.1007/978-981-13-1217-5_73} tested 576 instances across seven machine learning algorithms, including naive Bayes (NB), support vector machines (SVM), decision trees (DT), random forest (RF), bagging, adaptive boosting (AdaBoost), and linear regression (LR). Among these, AdaBoost and LR demonstrated the highest performance. Naik et al. \cite{Naik2023} further compared the effectiveness of RF, k-nearest neighbors (KNN), and SVM in classifying breast cancer histopathology images. By standardising image sizes and balancing the dataset, they enhanced the speed of model training and testing. Among these algorithms, RF achieved the best classification results. On the other hand, Labrada et al. \cite{9867097} assessed four machine learning methods; DT, SVM, KNN, and a Narrow Neural Network (NNN)—on the BreaKHis dataset \cite{7312934}, finding that NNN delivered the top performance.

Although classical machine learning methods have shown success in detecting and classifying breast cancer in histopathology images, they require complex feature extraction, which can be time-consuming and resource-intensive. In contrast, deep learning automatically extracts features, significantly improving both detection and classification performance \cite{Sharma2020, 7312934, basaad2024bert, Voon2022, Gupta2022, 10.3389/fonc.2024.1300997, 7312934}. Sharma et al. \cite{Sharma2020} compared traditional machine learning against deep learning models on the balanced BreaKHis dataset \cite{7312934}. They used Hu moments, color histograms, and Haralick textures for feature extraction in machine learning, while deep learning models included VGG16, VGG19, and ResNet50. The work found that the VGG16 model, paired with a linear SVM, provided the best results. On the other hand, Voon et al. \cite{Voon2022} applied seven different convolutional neural network (CNN) models: EfficientNetB0, EfficientNetV2B0, EfficientNetV2B0-21k, ResNetV1-50, ResNetV2-50, MobileNetV1, and MobileNetV2—on the FBCG dataset. Given the dataset’s imbalance, average performance metrics were used to evaluate results. The findings revealed that EfficientNetV2B0-21k performed best. Similarly, Gupta et al. \cite{Gupta2022} developed three distinct CNN models (ConvNet-A, ConvNet-B, and ConvNet-C) on a breast histopathology image dataset from Kaggle. These models, comprising 8, 9, and 19 layers respectively, were assessed using various performance metrics. The 19-layer model achieved the highest performance. Ramamoorthy et al. \cite{10.3389/fonc.2024.1300997} introduced the concept of SRGAN by merging traditional generative adversarial network (GAN) loss functions with components of efficient sub-pixel networks. Using the BreaKHis \cite{7312934} and IDC datasets \cite{dataset}, they applied Inception V3 and ResNet-50 (PFE-INC-RES) for block-level feature extraction, combined with transductive long short-term memory (TLSTM) to reduce false positives and improve classification accuracy.

\subsection{Model optimisation}
To enhance model performance and training efficiency, transfer learning has proven to be a powerful tool, allowing the extraction of valuable features from pre-trained models or datasets. This approach minimises the reliance on large volumes of labeled data, making it particularly useful for developing medical image analysis algorithms. For instance, Celik et al. \cite{CELIK2020232} used pre-trained models such as ResNet-50 and DenseNet-161 for IDC classification. Their results demonstrated that ResNet-50 achieved a balanced accuracy of 90.96\%, while DenseNet-161 performed slightly better with a balanced accuracy of 91.57\%. Similarly, Ghose et al. \cite{10.1007/978-981-33-6966-5_7} applied transfer learning by utilising feature maps from MobileNetV2, pre-trained on ImageNet, in conjunction with a deep convolutional neural network (DCNN) for IDC classification. They achieved a test accuracy of 98\% and a validation accuracy of 94\%, highlighting the effectiveness of this approach. In another study, Barsha et al. \cite{BARSHA2021104931} compared various post-transfer learning models, including ResNets, VGGs, DenseNets, and EfficientNets, in IDC detection experiments. Among these, DenseNet emerged as the top performer, achieving a balanced accuracy of 92.70\%. Several studies have also explored optimisations to CNN architectures to improve their performance. Chatterjee et al. \cite{8969037} introduced a deep residual convolutional network to analyse 7,500 breast histopathological images. This model simplifies the training of deep networks, enhancing efficiency, and achieved a remarkable accuracy of 99\%. Davoudi et al. \cite{Davoudi_Thulasiraman_2021} proposed the use of a genetic algorithm (GA) to optimise the weight parameters of CNN models. They conducted a comparative study of three optimisers: mini-batch gradient descent, Adam, and GA. By utilising GA, they addressed the issues of slow convergence and local optima commonly encountered with back propagation. The experimental results indicated that both the GA-optimised CNN and the Adam-optimised CNN achieved an accuracy of 85\%, demonstrating the effectiveness of the genetic algorithm in optimising model weights.

\subsection{Human-AI interaction}
Despite the significant advancements in artificial intelligence (AI), it still encounters challenges in addressing complex real-world problems, sometimes leading to unsatisfactory or incorrect outcomes. Weber et al. \cite{10.1145/3377325.3377509} explored the use of mchine learning-based model for image restoration, but found that machine learning alone could not produce satisfactory results. To overcome this limitation, they adopted a human-AI collaboration approach, where human intervention helped enhance the overall system performance.

Reverberi et al. \cite{Reverberi2022} demonstrated that hybrid decision-making systems, which combine both human expertise and AI, consistently outperform either humans or AI alone, particularly in medical decision systems. This synergy arises because the strengths of both humans and AI are utilised, capitalising on each party's unique capabilities. In human-AI collaboration, the role of humans is pivotal. Maadi et al. \cite{ijerph18042121} introduced the concept of human-in-the-loop machine learning (HILML), highlighting why, how, where, and who should be involved. Particularly in high-risk domains like medical diagnosis, they emphasised that the involvement of medical experts is crucial for enhancing AI performance, improving the quality and efficiency of outputs, and ensuring transparency and interpretability. However, they also pointed out that there is limited research on applying HILML in real-world medical systems, highlighting a gap that offers significant potential for future exploration. In a related work, Patil et al. \cite{PATIL2023100306} developed HistoROI, a lightweight classifier trained through a novel human-in-the-loop and active learning paradigm to classify histopathology whole-slide images. By integrating human feedback into the learning process, the model demonstrated strong generalisation ability. Boden et al. \cite{https://doi.org/10.1111/his.14356} applied a human-in-the-loop approach to digital image analysis in breast cancer diagnosis, and underscored that the human-in-the-loop approach improved error correction in specific cases. David et al. \cite{David} proposed a ``doctor-in-the-loop'' strategy based on the human-in-the-loop concept for testing whole-slide imaging data. Their experiment compared Bayesian hyperparameter optimisation with human-in-the-loop hyperparameter optimisation and found that the inclusion of human feedback improved the interpretability of the AI models, enhancing the overall diagnostic system performance. Additionally, Eduardo et al. \cite{Mosqueira-Rey2024} discussed active learning as a key example of HITL systems. In cases where data or labels are incomplete, the system identifies data that most urgently requires labelling, which is then annotated by human experts. The newly labelled data is subsequently used for model retraining, leading to enhanced data quality while reducing the overall labelling effort.

In summary, while AI has significantly enhanced accuracy and efficiency in diagnosing invasive ductal carcinoma (IDC), its role should primarily be seen as an assistive tool rather than a replacement for medical experts. The value of human expertise is indispensable, especially in optimising AI systems. Human-AI interaction should be bidirectional, where both entities provide mutual support and feedback to ensure high-quality decisions in clinical diagnoses.

\section{Methodology}
\label{sec:Methodology}
This section presents the proposed framework for the classification of histopathological breast cancer images. First, we describe the dataset considered for model evaluation in this work. Then, we outline the proposed framework, which consists of two main steps: pre-processing, and the structure of the EfficientNetV2S model within a deep learning framework. Lastly, we detail the integration of the human-in-the-loop (HITL) system.

\subsection{Dataset description}
\begin{figure}%[htbp]
\centerline{\includegraphics[width=0.8\textwidth]{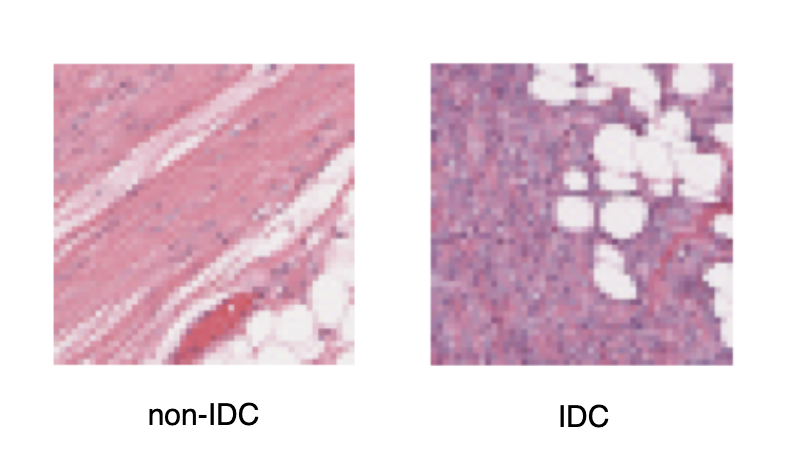}}
\caption{The left image shows healthy breast tissue, where muscle fibres are neatly and orderly arranged, indicating stable tissue architecture without obvious abnormalities. In contrast, the right image presents breast tissue affected by invasive ductal carcinoma (IDC). The dense acinar structures and their variations suggest the invasive nature of the cancer cells, with irregular arrangement and density of acini and ducts, typically associated with malignancy.}
\label{fig:Samples}
\end{figure}

Pathologists typically focus on identifying regions containing IDC to determine the invasiveness grade of the entire slide sample. In this work, we evaluate the performance of the proposed approach for the detection of IDC in histopathology images using the publicly available dataset \cite{dataset}. This dataset includes 162 whole-slide images (WSIs) containing breast cancer (BCa), all scanned at a $40\times$ magnification. From these WSIs, a total of 277,524 patches, each measuring $50\times 50$ pixels, are extracted. Among these patches, 198,738 are identified as non-invasive ductal carcinoma (IDC-negative), while 78,786 are confirmed as invasive ductal carcinoma (IDC-positive). Figure \ref{fig:Samples} presents examples of both non-IDC and IDC images.

\subsection{Data preprocessing}
To account for inter- and intra-patient variability, we preprocess the dataset images to standardise the data, making it more suitable for model training, thereby improving the model's accuracy, efficiency, and reliability. In this work, we consider five main preprocessing steps: data balancing, image normalisation, data augmentation, and data splitting. Below, we provide more details on each step.

\subsubsection{Data balancing}
In deep learning, the balance of the dataset directly impacts prediction results when dealing with classification problems. When the number of different data types in the dataset is uneven, it causes data skew, meaning the model's predictions will be biased towards one category \cite{asi5050087}. Data balancing ensures that the number of samples for each category in the dataset is roughly equal. In this work, we randomly select 7,000 samples from both the IDC-negative and IDC-positive categories for training.

\subsubsection{Normalisation}
Normalisation can help optimisation algorithms, such as gradient descent, converge faster. Unnormalised data may lead to uneven convergence rates in model training, as different features may have different value ranges and units. This can cause some features to dominate the gradient calculation, affecting the overall learning process. In this work, we normalise the samples by scaling them to the range [0, 1]. Each histopathology image is then transformed into a single vector form of $7500\times 1$ \cite{Gupta2022}, a process commonly referred to as flattening, which converts the multidimensional array of images $(50\times 50 \times 3)$ into a long one-dimensional array.

\subsubsection{Data augmentation}
For a dataset comprising 198,738 IDC-negative and 78,786 IDC-positive images, addressing class imbalance is essential, with data augmentation strategies playing a crucial role in improving model performance. By generating diverse variations, such as flips, rotations, and scaling, the sample space of the minority class is expanded, fostering a more balanced data distribution and enabling the model to better capture minority-class features. Moreover, combining data augmentation with small-scale datasets allows for rapid validation of the model's response to these strategies, streamlining debugging and performance evaluation. Extracting data subsets further reduces computational time and storage requirements, making it an efficient solution when training resources, such as GPU memory or processing power, are limited. These techniques involve various transformations of the images, effectively increasing the diversity of the dataset \cite{Gupta2022, Koshy2022, s22030876}. In this work, the ImageDataGenerator method provided by the Keras deep learning library is used \cite{s20164373}. This method generates two transformed samples from each original image, thereby expanding and diversifying the dataset \cite{chaudhury_2023_deep, 10.1007/978-981-16-8774-7_53}. The following transformations are considered:

\begin{itemize}
\item {Rotation}: The image is rotated randomly at load time by an angle between 0 and 30 degrees.
\item {Random Horizontal Shifts}: Images are moved horizontally by up to 20\% of the image width.
\item {Random Vertical Shifts}: Images are moved vertically by up to 20\% of the image height.
\item {Shear Transformations}: One part of the image is tilted with a shear strength of 0.2 while keeping another part unchanged.
\item {Random Zoom}: The image is resized randomly at load time by a factor of $1 \pm 0.2$ (from 80\% to 120\%).
\end{itemize}

\begin{figure*}%[htbp]
\centerline{\includegraphics[width=0.88\textwidth]{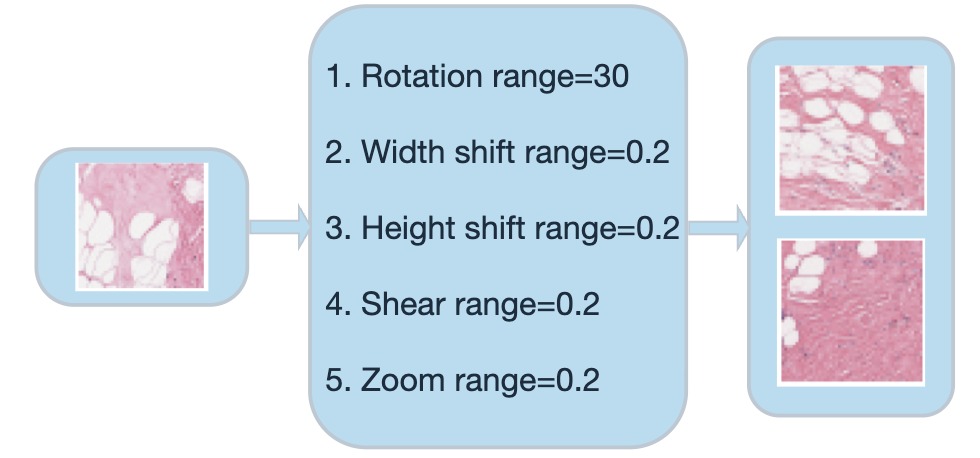}}
\caption{The figure illustrates the application of image augmentation techniques in processing histopathology images. By applying various transformations to an original microscope image, such as rotation (up to 30 degrees), width shift (up to 20\%), height shift (up to 20\%), shear transformation (up to 20\%), and zoom (up to 20\%), two new images can be generated.}
\label{fig:Augm}
\end{figure*}

Figure \ref{fig:Augm} shows a schematic diagram of the five data augmentation techniques considered in this work to increase the number of training data samples.

\subsubsection{Data splitting}
In this work, a total of 28,000 processed data samples are divided into training and testing sets. The training set, comprising 70\% of the data (19,600 images), is used for model training, while the testing set, comprising 30\% of the data (8,400 images), is used to evaluate the model's generalisation ability.

\subsection{Deep learning model}
The paper proposes a framework based on transfer learning, leveraging the efficient EfficientNetV2 model \cite{electronics12061342}. EfficientNet, originally introduced by Tan et al. \cite{tian}, is an advanced architecture built on convolutional neural networks. Its core innovation lies in a systematic model scaling method that optimizes the network's depth, width, and input image resolution simultaneously, achieving an optimal balance between performance and computational resources. This design significantly enhances the model's efficiency and accuracy across various computational tasks. Building on this, EfficientNetV2 further improves the architecture by introducing hybrid convolutional modules and optimizing the training pipeline, resulting in faster training, fewer parameters, and maintained or improved classification accuracy across multiple tasks \cite{Abioye_Evwiekpaefe_Awujoola_2024}.

In this study, we developed a deep convolutional neural network model based on EfficientNetV2, pretrained on the ImageNet dataset. ImageNet, a widely used large-scale image dataset, provides rich semantic information and diverse feature representations, offering a comprehensive knowledge base for the model. Through transfer learning, the pretrained weights are utilized for model initialization, effectively leveraging the knowledge learned from the large-scale dataset. This approach significantly improves the model's efficiency, accuracy, and generalization ability \cite{Souza2024, chaudhury_2023_deep, 9332405}. Furthermore, transfer learning reduces the dependency on extensive labeled data and greatly decreases training time and computational costs.

\begin{figure}%[htbp]
\centerline{\includegraphics[width=0.99\textwidth]{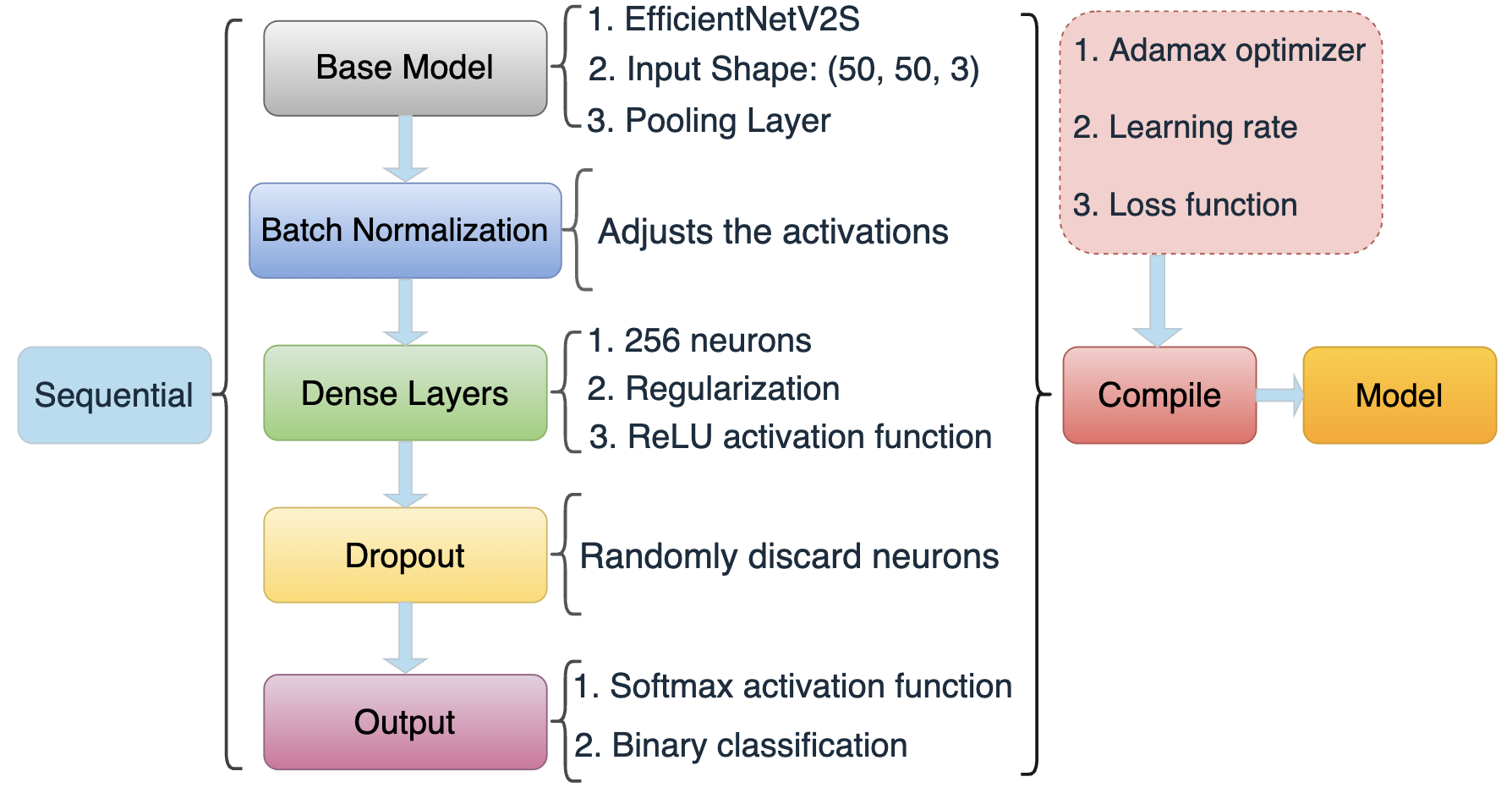}}
\caption{The figure shows a binary classification neural network model based on the EfficientNetV2S architecture, processing $50 \times 50$ pixel images. The model includes pooling layers, batch normalisation, a dense layer with 256 neurons, dropout layers to reduce overfitting, and an output layer using the softmax function. The model is compiled using the Adamax optimiser with a defined learning rate and loss function.}
\label{fig:ProposedApproach}
\end{figure}

The proposed model design is illustrated in Figure \ref{fig:ProposedApproach}. In its implementation, the model utilizes EfficientNetV2 for robust feature extraction, combined with a carefully designed subsequent layer structure, delivering exceptional performance in image classification tasks. The multi-resolution processing capability and efficient architecture of EfficientNetV2 ensure high performance when handling high-resolution images while maintaining low computational resource consumption.

The model architecture begins by loading the pre-trained EfficientNetV2S without the top layer, leveraging its powerful feature extraction capabilities. Custom layers are then added to further process features and perform classification tasks. Specifically, the top fully connected layer is removed, and max pooling is applied at the end of the model to extract key information from the feature maps. Subsequently, a batch normalisation layer is added to stabilise and accelerate the training process by normalising the activations, thus enhancing the model's learning efficiency. The batch normalisation is computed as follows
\begin{equation}
\begin{aligned}
\hat{x}^{(k)} &= \frac{x^{(k)} - \mu_B}{\sqrt{\sigma_B^2 + \epsilon}}, \\
y^{(k)} &= \gamma \hat{x}^{(k)} + \beta,
\end{aligned}
\end{equation}
where $\mu_B$ and $\sigma_B^2$ are the mean and variance of the batch, $\epsilon$ is a small constant for numerical stability, and $\gamma$ and $\beta$ are learnable parameters. Following this, a fully connected layer with 256 units and the ReLU activation function is added, which is defined as
\begin{equation}
f(x) = \max(0, x).
\end{equation}

To mitigate overfitting and improve model generalisation, $\ell_1$ and $\ell_2$ regularisations are applied to this layer, defined as:
\begin{equation}
\begin{aligned}
L_1 &= \lambda_1 \sum |w|, \\
L_2 &= \lambda_2 \sum w^2,
\end{aligned}
\end{equation}
where $\lambda_1$ and $\lambda_2$ are the regularisation coefficients. This layer also includes Dropout to randomly drop some nodes during training, enhancing the model's adaptability and robustness to new data. 

Finally, the model employs an output layer with a softmax activation function to generate the predicted probabilities for the two classes, suitable for the binary classification task in this work. The softmax activation function, which converts logits into probabilities, is given by
\begin{equation}
\sigma(z_i) = \frac{e^{z_i}}{\sum_{j} e^{z_j}},
\end{equation}
where $z_i$ are the logits. During the model compilation phase, the Adamax optimiser is used with a defined learning rate to ensure fine-tuning of parameters near the optimal solution. The loss function is categorical cross-entropy, which is commonly used for classification tasks, and the primary performance metric for the model is classification accuracy.

\subsection{Human-AI collaboration system}
After training a high-performance model using the aforementioned methods, we implemented a Human-in-the-Loop (HITL) interactive system to further optimize the model's performance. In this system, we designed an interactive interface that allows users to invoke the trained model to predict on unseen images and display the results intuitively. Users are not merely observers but key participants: they evaluate the model's predictions, and if misclassified samples are identified, users can correct the labels and add these corrected samples back into the training dataset. The model is then retrained using the expanded dataset. This process fully leverages human intervention, significantly enhancing the model's performance and adaptability.

This method shares similarities with semi-supervised learning \cite{https://doi.org/10.1155/2020/8826568}, but it has a notable difference: it incorporates pseudo-labeled data derived from prediction errors and other studies while integrating human expertise. This approach offers unique advantages in improving data quality and diversity.

\begin{figure}%[htbp]
\centerline{\includegraphics[width=0.88\textwidth]{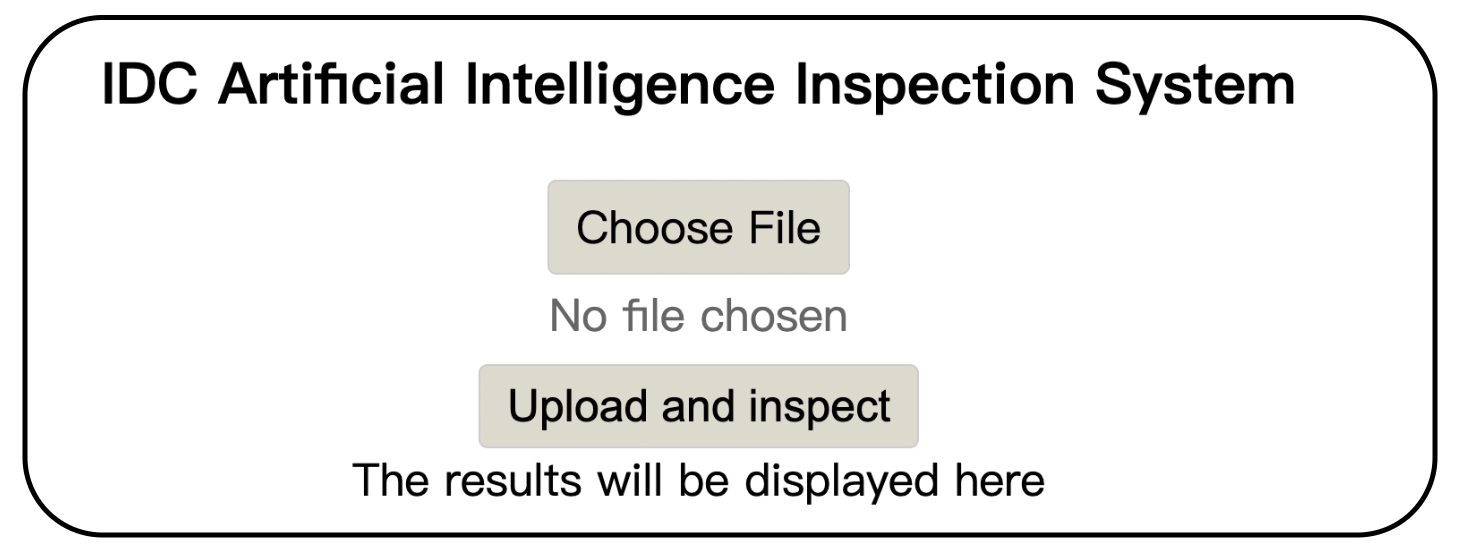}}
\caption{The figure shows the interactive interface design. The user can select a file by clicking the "Choose File" button and then click the "Upload and Inspect" button to upload and inspect the file. The results will be displayed below.}
\label{fig:GUI}
\end{figure}

To enhance usability, we designed a user-friendly interface that allows users to easily input images, click a prediction button, and quickly view the model's results. The backend system is developed in Python, while the frontend interface is built using an HTML framework, as illustrated in Figure \ref{fig:GUI}. The core of the HITL system is to utilize user-corrected misclassified samples to iteratively improve the model through the collaboration of human and artificial intelligence. By correcting erroneous labels, users not only refine the dataset but also introduce novel features, enabling the model to learn from previously unrecognized patterns.

\begin{figure*}%[htbp]
\centerline{\includegraphics[width=0.99\textwidth]{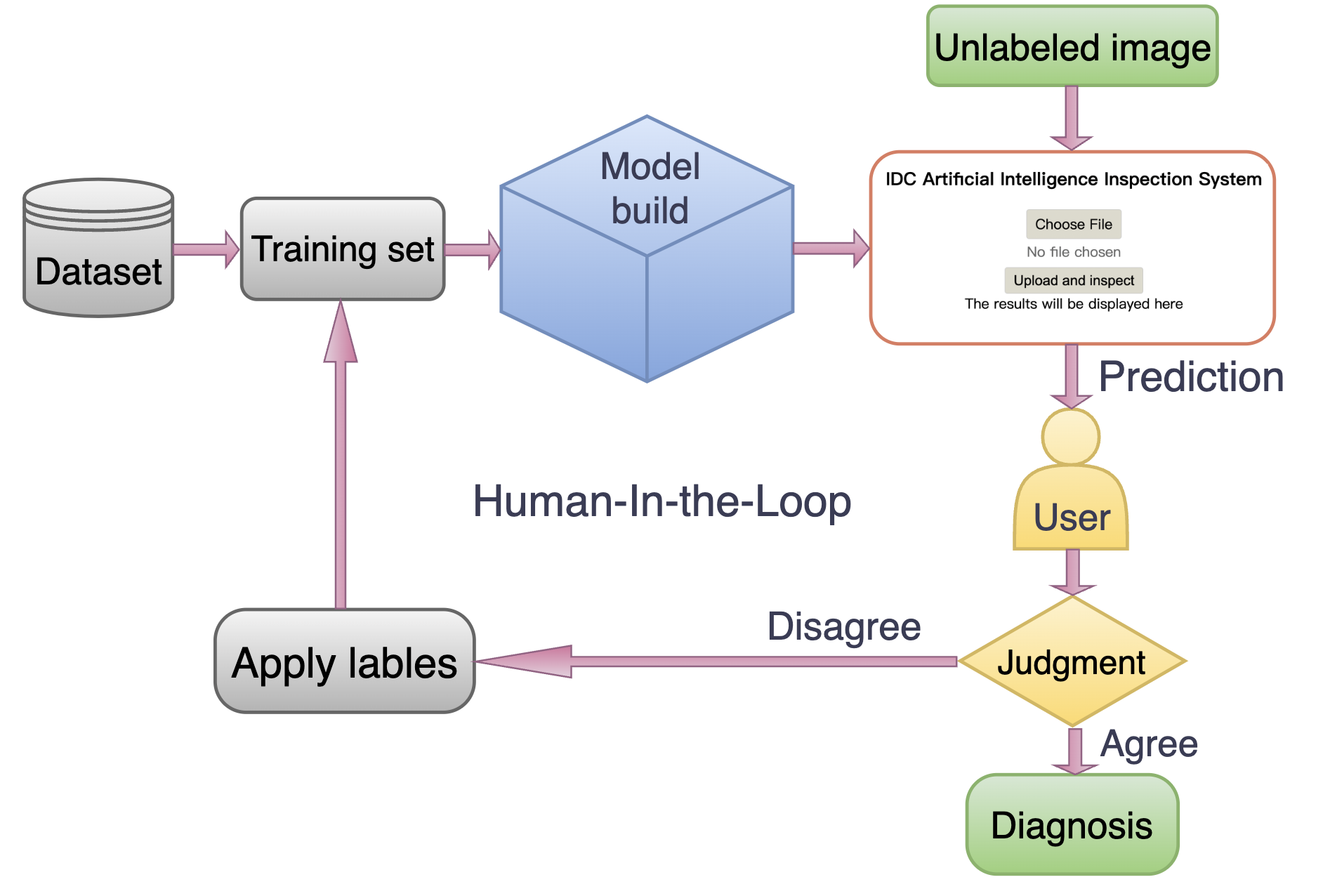}}
\caption{The figure illustrates an interactive model of human-in-the-loop (HITL) for medical AI systems. The system builds a model from the dataset and uses this model to make predictions on unlabeled images. Users review the AI's predictions according to their professional judgement, and if they agree, it becomes the final diagnosis. If they disagree, they annotate the data and provide feedback, and the model is retrained to optimise performance.}
\label{fig:HitlFlowchart}
\end{figure*}

The entire workflow, as shown in Figure \ref{fig:HitlFlowchart}, consists of two key phases:
\begin{enumerate}
\item Model Training and Testing: Selecting the optimal version by comparing performance metrics.
\item Validation of the HITL Approach: Users evaluate and correct model predictions through the interactive interface, validating the effectiveness of this approach in improving model performance and adaptability.
\end{enumerate}
Through this human-machine collaboration, the HITL system not only enhances classification accuracy but also demonstrates the irreplaceable role of human expertise in feature discovery and problem correction during the learning process.

\subsubsection{Model training}
\begin{figure*}%[htbp]
\centerline{\includegraphics[width=0.99\textwidth]{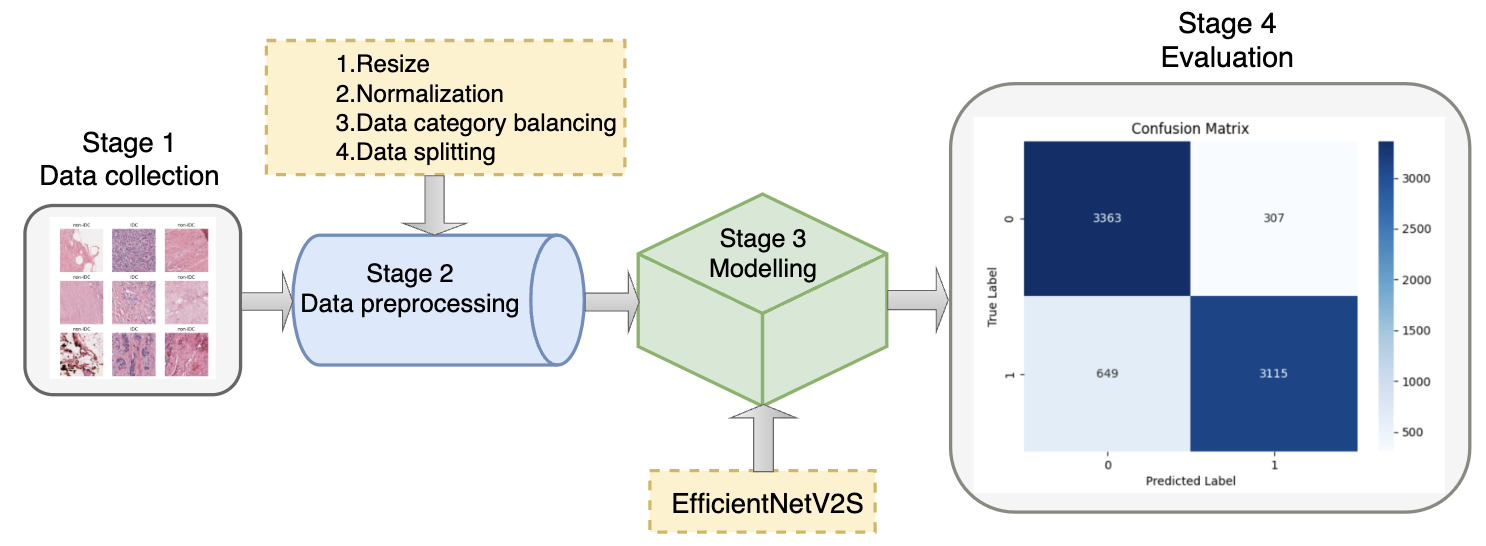}}
\caption{The figure illustrates the model training flow, divided into four stages. The first phase is data collection, gathering the required medical images. The second stage involves data pre-processing, including image resizing, normalisation, data class balancing, and dataset splitting. The third stage involves modelling with the EfficientNetV2S model. The final stage evaluates model performance using the confusion matrix.}
\label{fig:TrainingFlowchart}
\end{figure*}

From the dataset, 7,000 samples each from negative and positive cases were selected. After pre-processing and data augmentation, the number of negative and positive samples increased to 14,000 each, totalling 28,000. Of these, 70\% will be used for training, and 30\% for testing. The model begins by loading EfficientNetV2S without the top layer, initialised with ImageNet weights, and applies max pooling to the output. In the batch normalisation layer, the normalisation axis is set to -1, the decay factor to 0.99, and $\epsilon$ to 0.001. The fully connected layer has 256 neurons with a ReLU activation function, and L1 and L2 regularisation are set to 0.006. The dropout rate is set to 0.45, meaning 45\% of the neurons are randomly dropped during each training iteration. The output layer is a fully connected layer with two units, using the softmax function to output predicted classification probabilities. During compilation, the model uses the Adamax optimiser with a learning rate of 0.0001, the loss function is categorical cross-entropy, and the evaluation metric is accuracy. The number of epochs is set to 100. The model with the best performance parameters will be saved. Figure \ref{fig:TrainingFlowchart} shows the flowchart of the model training process.

\subsubsection{Verifying the human-in-the-Loop system}
\begin{figure*}%[htbp]
\centerline{\includegraphics[width=0.99\textwidth]{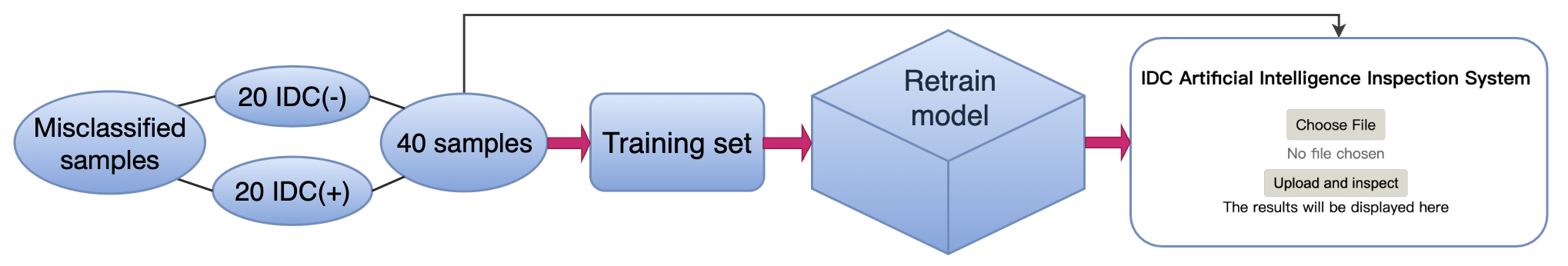}}
\caption{This diagram illustrates the retraining process of the AI model. Initially, 20 negative and 20 positive samples are taken from each misclassified sample, resulting in 40 samples. These are added to the training set to retrain the model. Finally, the new model is used to predict 40 new samples to verify its effectiveness.}
\label{fig:ValidationFlowchart}
\end{figure*}

To validate the effectiveness of the proposed method, we utilized the newly trained model to predict and analyze misclassified samples from previous training sessions. The detailed flow of the validation experiment is shown in Figure \ref{fig:ValidationFlowchart}. The experiment design and operational steps are as follows.
\begin{itemize}
\item We selected misclassified samples from the test set of the initial model, including 20 negative samples (actual negatives misclassified as positives) and 20 positive samples (actual positives misclassified as negatives), totaling 40 images. As these samples were misclassified by the initial model, their initial classification accuracy is expected to be 0\%.
\item The 40 misclassified samples were added to the original training set to create a new extended training set. The model was then retrained on this extended dataset to observe whether it could correct the misclassification of these samples through incremental learning.
\item The retrained model was used to predict the classifications of the 40 misclassified samples, and the classification results were recorded. By comparing these results with the true labels, the new classification accuracy was calculated to assess the model's improvement in handling these misclassified samples.
\item To ensure objectivity and minimize the impact of random sample selection, four independent experiments were conducted with different sets of misclassified samples. For each experiment, new misclassified samples were selected, an extended training set was constructed, the model was retrained, and accuracy was tested. 
\end{itemize}

\section{Experimental Results}
\label{sec:Results}
This section is divided into three parts. The first part introduces the evaluation metrics used to assess the model's performance. The framework evaluation is further subdivided into two approaches: testing the AI-based system independently and testing with the integration of the Human-in-the-Loop (HITL) system. The second part showcases the model's predictive capabilities, enhanced through the incorporation of misclassified images via the HITL system.

\subsection{Evaluation criteria}
The classification performance of our proposed model for histopathology breast cancer images is assessed using the confusion matrix, also known as the error matrix or contingency table. This matrix contains four key terms: True Positive (TP), False Positive (FP), False Negative (FN), and True Negative (TN). In our context, TP refers to images correctly classified as carcinoma, FP represents non-carcinoma images mistakenly classified as carcinoma, FN denotes carcinoma images misclassified as non-carcinoma, and TN refers to non-carcinoma images correctly classified. The model’s performance is evaluated on the testing set using four metrics derived from the confusion matrix; accuracy, sensitivity, specificity, precision, and F1-score, and can be calculated as follows.

\begin{equation}
\text{Accuracy} = \frac{\text{TP} + \text{TN}}{\text{TP} + \text{TN} + \text{FP} + \text{FN}},
\end{equation}

\begin{equation}
\text{Sensitivity} = \frac{\text{TP}}{\text{TP} + \text{FN}},
\end{equation}

\begin{equation}
\text{Specificity} = \frac{\text{TN}}{\text{TN} + \text{FP}},
\end{equation}

\begin{equation}
\text{Precision} = \frac{\text{TP}}{\text{TP} + \text{FP}},
\end{equation}

\begin{equation}
\text{F1-score} = 2 \times \frac{\text{Precision} \times \text{Recall}}{\text{Precision} + \text{Recall}},
\end{equation}
where accuracy is the proportion of correct samples to total samples, sensitivity is the percentage of all IDC cases correctly identified as IDC, specificity is the percentage of non-IDC cases correctly identified, precision is the proportion of total samples classified correctly, and F1-score is a common metric used to measure the accuracy of a classification model.

\subsection{Results of the AI system}
\begin{figure}%[htbp]
\centerline{\includegraphics[width=0.65\textwidth]{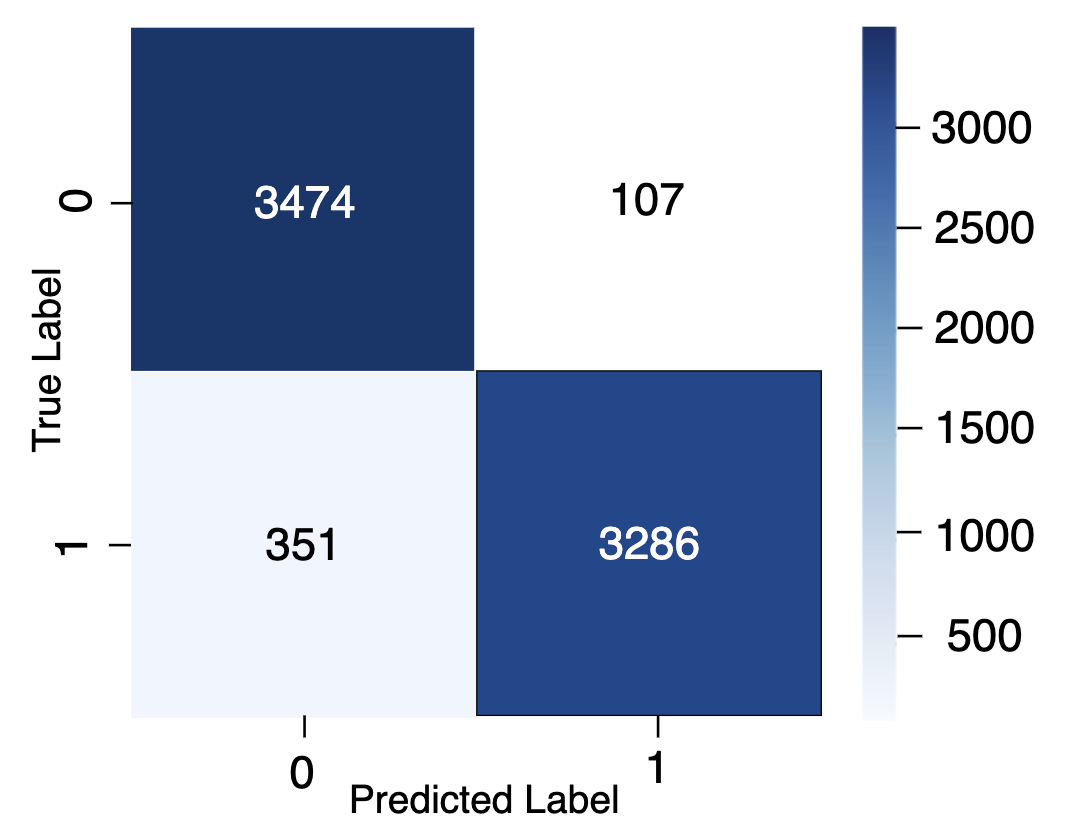}}
\caption{The confusion matrix for the proposed model. TN is 3474, FP is 107, FN is 351, and TP is 3286, which can be used to compute the performance metrics.}
\label{fig:Conf}
\end{figure}

\begin{figure}%[htbp]
\centerline{\includegraphics[width=0.99\textwidth]{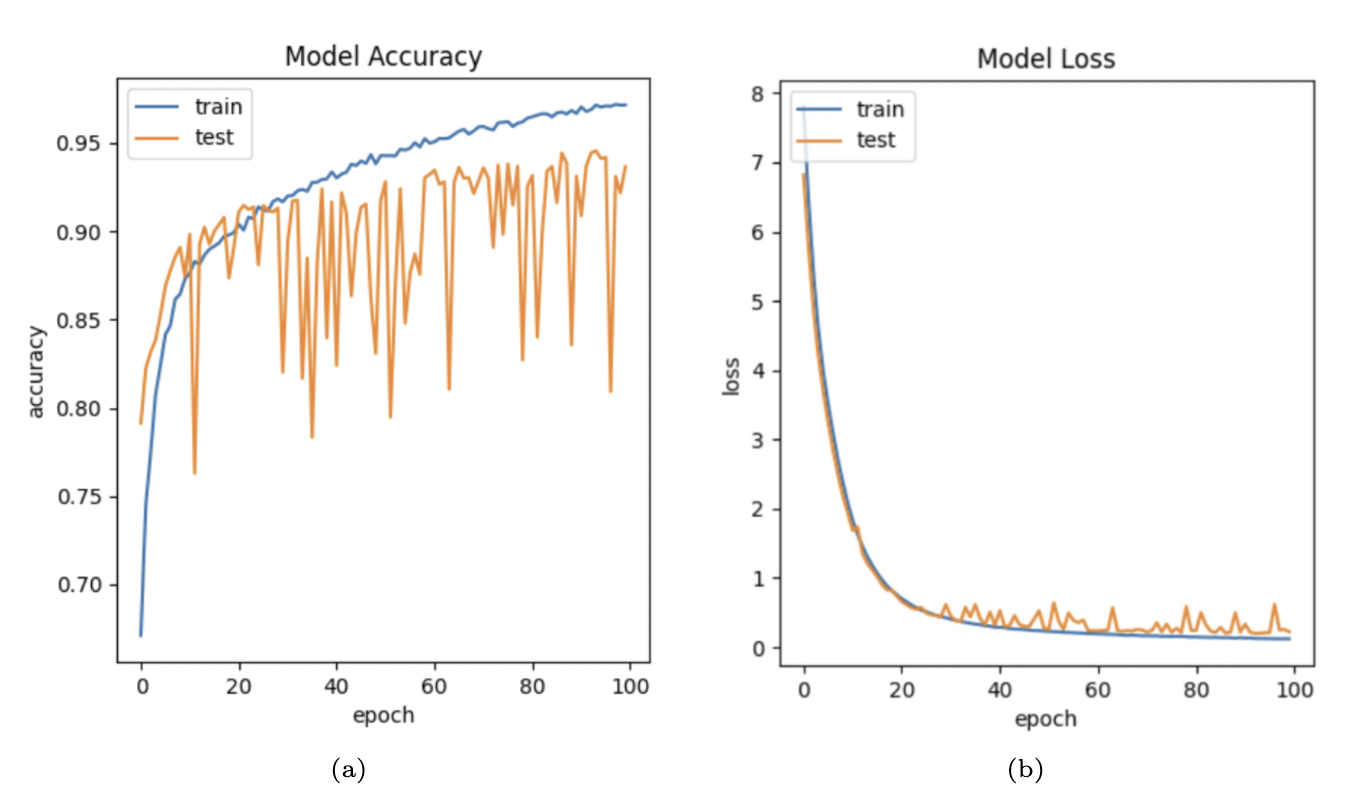}}
\caption{(a) Model accuracy over epochs for training and test sets. The blue line indicates training accuracy, which rises and stabilises around 97\%, while the orange line shows test accuracy, which stabilises around 95\% after 100 epochs. This indicates that the model performs well on both training and test sets without overfitting. (b) Training and test loss as a function of epochs. The blue line represents training loss, and the orange line represents test loss. Both losses decrease and stabilise at low values, indicating a good fit on the training and test data.}
\label{fig:test}
\end{figure}

Figure \ref{fig:Conf} shows the confusion matrix of the model evaluated using the dataset described in the previous section. The calculated performance metrics are: Accuracy: 93.65\%, Sensitivity (Recall): 90.35\%, Specificity: 97.01\%, Precision: 96.85\%, and F1-score: 93.49\%. On the other hand, Figure \ref{fig:test} present the accuracy and loss graphs of the model. These graphs are essential for evaluating and fine-tuning deep learning models, offering a visual representation of performance trends during training. The x-axis represents the epochs, while the y-axis denotes accuracy and loss values, respectively. The blue line corresponds to the training set data, and the orange line represents the testing set data. Typically, a gradual increase in accuracy and a decrease in loss values are desirable as training progresses. In particular, Figure \ref{fig:test}(a) shows that training accuracy (blue line) rises and stabilises at around 97\%, and the test accuracy (orange line) stabilises at around 95\% after 100 epochs. This indicates that the model performs well on both training and test sets without overfitting. On the other hand, Figure \ref{fig:test}(b) shows training and test loss as a function of epochs. Both training and testing losses decrease and stabilise at low values, indicating a good fit on the training and test data.

\subsection{Comparison with existing methods}
In this work, we compare the performance of the proposed approach with the latest frameworks evaluated using the same dataset \cite{dataset}. In particular, we consider: Celik et al. \cite{CELIK2020232}, Barsha et al. \cite{BARSHA2021104931}, Chaudhury et al. \cite{chaudhury_2023_deep}, Wang et al. \cite{electronics11172767}, Romero et al. \cite{8759410}, Gupta et al. \cite{Gupta2022}, and Roy et al. \cite{s21113628} for comparison. Table \ref{tab:example} provides the evaluated metrics. The comparison is based on accuracy, sensitivity, specificity, precision, and F1-score values. The proposed model achieves the highest detection accuracy of 93.65\%. It also records the highest precision (96. 85\%) and specificity (97. 01\%), indicating its effectiveness in correctly identifying true negatives and reducing false positives, which is crucial in medical diagnostics. While some existing methods, such as ResNet-50 demonstrates marginally better sensitivity, and DenseNet-121, demonstrates marginally better F1-score, the proposed approach strikes an optimal balance between accuracy and specificity, making it a more reliable choice for IDC classification. 

\renewcommand{\arraystretch}{2}
\begin{table}%[ht]
\centering
\caption{Comparison with state-of-the-art methods on the same dataset. Acc stands for Accuracy, Sen for Sensitivity, Spec for Specificity, Prec for Precision, and F1 for F1-score. All values are represented as percentages (\%). Best results are in bold, while second best are underlined.}
\label{tab:example}
\begin{tabular}{@{}cccccccc@{}}
\toprule
                                        & \textbf{Model}        & \textbf{Acc}   & \textbf{Sen}   & \textbf{Prec}  & \textbf{Spec}  & \textbf{F1}    \\ \midrule
Proposed                         & EfficientNetV2S         & \textbf{93.65} & {90.35} & \textbf{96.85} & \textbf{97.01} & {93.49} \\
%\cite{10548784}  & EfficientNetV2S & \textbf{95.99} & 85.28 & 88.27 & \textbf{97.94} & 86.75 \\
\cite{CELIK2020232}       & DenseNet-161 & 91.20 & 89.59 & \underline{95.34} & 93.56 & 92.38 \\
\cite{CELIK2020232}       & ResNet-50    & 91.96 & \textbf{93.64} & 94.58 & 88.28 & \underline{94.11} \\
\cite{BARSHA2021104931}  & DenseNet-121 & {92.47} &    ---   &    ---  &    ---   & \textbf{94.50} \\
\cite{chaudhury_2023_deep} & SqueezeNet & 89.49 & 78.70 & 83.46 & 93.78 & 81.01 \\
\cite{electronics11172767} & CNN-GRU      & 86.21 & 85.71 & 85.90 & 84.51 & 86.00 \\
\cite{8759410}           & Inception    & 88.50 & 90.00 & 87.38 & 87.00 & 88.67 \\
\cite{Gupta2022}          & ConvNet-C    & 88.70 & \underline{92.60} &   ---    &  ---     &   ---    \\
\cite{s21113628}          & CatBoost     & 92.50 & 88.80 & 93.40 &  ---     & 90.70 \\ \bottomrule
\end{tabular}
\end{table}

\subsection{Integration of the Human-in-the-Loop system}
Through four sets of experiments, we validated the critical role of the Human-in-the-Loop (HITL) approach in improving model performance. In each experiment, humans actively intervened by selecting misclassified images from the model's errors and incorporating them into an enhanced training dataset. The goal was to evaluate whether human-selected misclassified samples could effectively boost the model's performance. Table \ref{tab:my-table} summarizes the accuracy changes before and after integrating the HITL system.

The experimental results demonstrate that the HITL approach significantly improved the model's classification ability. Initially, the model failed to correctly classify any images in each group, with a baseline accuracy of 0\%. However, after human intervention in selecting and re-annotating misclassified samples and adding them to the training dataset, the model's performance improved substantially.

Specifically, in Group 1, the accuracy increased to 70\% after incorporating human-selected misclassified images, with 28 out of 40 misclassified images correctly reclassified. In Group 2, thanks to further optimization in the selection of samples by humans, the improvement was even more pronounced, with accuracy rising to 85\%, correctly classifying 34 out of 40 images. Similarly, the results of Group 3 and Group 4 also highlighted the effectiveness of the HITL approach, achieving accuracies of 77.5\% and 75\%, respectively.

\begin{table}%[ht]
\centering
\caption{Accuracy of the model before and after adding misclassified images to the training set in the Human-in-the-Loop experiment.}
\label{tab:my-table}
\begin{tabular}{cccccc}
\hline
\multicolumn{2}{c|}{} & \multicolumn{2}{c|}{\textbf{Before}} & \multicolumn{2}{c}{\textbf{After}} \\ \hline
\textbf{Group}     & \textbf{Total}     & \textbf{Correct}      & \textbf{Accuracy}     & \textbf{Correct}     & \textbf{Accuracy}    \\ \hline
1         & 40        & 0            & 0\%          & 28          & 70\%        \\
2         & 40        & 0            & 0\%          & 34          & 85\%        \\
3         & 40        & 0            & 0\%          & 31          & 77.5\%      \\
4         & 40        & 0            & 0\%          & 30          & 75\%        \\ \hline
\end{tabular}
\end{table}

\section{Discussion}
\label{sec:Discussion}
This work developed a human-AI collaboration system to enhance the detection accuracy of invasive ductal carcinoma (IDC) in histopathology images. The discussion focuses on two key aspects: the performance of the EfficientNetV2S model and the impact of integrating a Human-in-the-Loop (HITL) approach into the diagnostic process. 

The EfficientNetV2S model demonstrated superior performance in IDC classification, achieving an accuracy of 93.65\%, sensitivity of 90.35\%, specificity of 97.01\%, precision of 96.85\%, and an F1-score of 93.49\%. These results underscore the model's effectiveness in distinguishing between IDC-positive and IDC-negative cases. The high specificity and precision are particularly noteworthy, as they reflect the model's ability to minimise false positives, which is crucial in medical diagnostics where incorrect diagnoses can lead to unnecessary anxiety and additional procedures \cite{s20164373, rai2024, Sohail2021}. Comparative analysis with other state-of-the-art models, such as DenseNet-161 and ResNet-50, reveals that EfficientNetV2S provides a balanced approach to computational efficiency and accuracy. DenseNet-161 achieved an accuracy of 91.20\% and ResNet-50 reached 91.96\%, while EfficientNetV2S outperformed these models by leveraging a streamlined architecture that enhances both training speed and parameter efficiency. This finding aligns with the broader literature on deep learning in medical imaging, where EfficientNet models are recognised for their ability to optimally balance depth, width, and resolution \cite{CELIK2020232, BARSHA2021104931, chaudhury_2023_deep, electronics11172767, 8759410, s21113628}.

On the other hand, the HITL approach was evaluated by incorporating misclassified images into the training set. The results indicate significant improvements in the model's ability to classify previously misclassified images correctly, demonstrating that adding erroneous data to the training set can enhance model performance and generalisation. This approach, akin to semi-supervised learning, not only corrects previous errors but also improves the model's robustness through iterative training. The integration of a HITL system substantially improved model performance by incorporating human expertise into the iterative training process. By correcting misclassifications and enriching the training dataset with more accurate labels, the HITL system enhanced the model's accuracy, with correction rates ranging from 70\% to 85\%. This iterative feedback loop aligns with active learning principles, where the model learns from the most informative data points \cite{mosqueira2023}. Active learning, particularly when combined with human oversight, has been shown to improve model robustness in complex medical diagnostics \cite{romero2019}. Moreover, the HITL system also addresses key limitations of AI in medical diagnostics, such as interpretability. By allowing human experts to intervene and correct AI predictions, the system ensures that the model's outputs align more closely with clinical expectations. This collaboration fosters trust in the system by combining the computational power of AI with the nuanced judgment of medical professionals.

While the benefits of the proposed approach are clear, a few limitations remain. The fixed image size of $50 \times 50$ pixels may have constrained the model's ability to capture finer details in histopathology images. Future work could explore the impact of larger image sizes, such as $250 \times 250$ pixels. However, this might come at a higher computation complexity. On the other hand, the effect of varying image magnifications can be investigated, as this could influence model accuracy. Future work could also consider expanding the HITL system to include other feedback mechanisms, such as prediction maps or active learning strategies, which could further enhance the system's adaptability to new data and improve its interpretability. 

\section{Conclusions}
\label{sec:Conclusions}
This work presented a novel Human-AI collaboration system designed to enhance the detection of invasive ductal carcinoma in histopathology images. By integrating the EfficientNetV2S model with a Human-in-the-Loop (HITL) approach, we demonstrated substantial improvements in both model accuracy and generalisation capabilities. The EfficientNetV2S model itself achieved state-of-the-art results. Moreover, the integration of the HITL system was shown to be highly effective in further improving model performance. Through the iterative process of incorporating misclassified images into the training dataset, we observed significant performance gains across four experimental groups. This demonstrates the potential of leveraging human expertise to correct model errors and enhance the system’s robustness, especially in challenging cases where the model initially struggled. Overall, this work highlights the importance of combining advanced deep learning techniques with human expertise to improve the reliability and interpretability of AI systems in medical diagnostics. The HITL framework offers a promising avenue for future research, particularly in refining model outputs and ensuring more accurate diagnoses in clinical practice.\\

\noindent\textbf{Conflict of interest}\\
The authors declare no conflict of interests.

\bibliography{Final-Han-Springer}% common bib file

\end{document}